\documentclass[aps,pre,preprint,showpacs]{revtex4}
\usepackage{graphicx,psfrag,subfigure}
\usepackage{amsmath,amssymb,revsymb}
\usepackage{color}

\def\bU{{\bf U}}

\def\br{{\bf r}}

\def\bB{{\bf B}}
\def\bb{{\bf b}}

\begin{document}

\title{Fluctuation of magnetic induction in von K\'arm\'an swirling flows}
\author{Romain VOLK, Philippe ODIER, Jean-Fran\c{c}ois~PINTON}
\address{Laboratoire de Physique - UMR 5672,\\
CNRS \& Ecole Normale Sup\'erieure,
46 All\'ee d'Italie, F-69007 Lyon, France}

\begin{abstract}
Studies of magnetic induction in von K\'arm\'an swirling flows have so far linked the time-averaged induced magnetic field to the structure of the mean flow. They have evidenced  the Omega and Parker mechanism generated respectively by the flow differential rotation and helicity, which underly the Duddley and James~\cite{DuddleyJames} dynamos. Using an array of Hall probes  we study here the dynamical regime. In the experimental flow, turbulence is fully developed and large fluctuations are observed in the magnetic induction processes. We find that the large scale turbulent fluctuations have different characteristics when induction results from the differential rotation or from the dynamics of the stagnation point in the mid plane of the von K\'arm\'an flow. Symmetry considerations indicate that the dynamical flow spends half of its time away from the time-averaged  structure. The consequences of these observations for dynamo experiments are discussed.
\end{abstract}

\pacs{47.65.+a,47.27.-i,91.25.Cw}
\date{ \today}

\maketitle

\section{Introduction}
The von K\'arm\'an flows generated in the gap between counter-rotating impellers has been considered by several groups as a possible candidate~\cite{Maryland,Wisconsin,VKSPoF} in the search for a laboratory demonstration of a homogeneous fluid dynamo that would be less constrained than the Riga and Karlsruhe designs~\cite{Riga,Karlsruhe}. In these experiments, liquid sodium is often used because its electrical conductivity is high (half that of copper, at $120^{\circ}$C), while its density remains low, of the order of that for water. Its magnetic diffusivity, as for all metals,  is many orders of magnitude larger than its hydrodynamic viscosity. The flow Reynolds number needs to be very large in order for non-linearities to develop in the magnetic induction. As a consequence, the hydrodynamic flow is very turbulent, and many questions arise concerning the influence of turbulence on the bifurcation threshold and the dynamics in an eventual saturated regime. 

This problem is very complex, and many studies so far have focused on the dynamo capacity of the average flow engineered in the von K\'arm\'an (VK) geometry. Here, `average' has the meaning of `time-average'. A time-averaged flow field $\langle \bU\rangle (\br)$ is derived from measurements in water prototypes (for convenience and because velocimetry methods are scarce in opaque, high temperature flows of
 liquid metals~\cite{DOPNataf,DOPGerbeth}) as:
\begin{equation}
\langle \bU \rangle (\br) = \frac{1}{T} \int_0^{T} \bU(\br, t) dt \ ,
\label{eqMoy}
\end{equation}
where $T$ is a time much longer than the one characteristic of the forcing of the flow (estimated, for instance, as the period of rotation of the driving impellers). This stationary flow profile --  no longer a solution of the Navier-Stokes equation -- is then usually inserted into a kinematic numerical solver, in which the induction equation
\begin{equation}
\partial_t \bB = \nabla \times \left( \langle \bU \rangle \times \bB \right)  + \lambda \Delta \bB \ ,
\label{eqB}
\end{equation}
($\lambda$ is the magnetic diffusivity) is solved with the velocity field kept constant in time~\cite{Marie,ForestKinematic,Ravelet}. These studies have shown the possibility of dynamo action in the average VK flows, and the underlying induction processes have been measured in sodium and Gallium flows~\cite{VKSPoF,bourgoinvolk}  and analyzed in details~\cite{bourgoinPoF,MarieThesis,BourgoinThesis}. It showed that the helicity and differential rotation present in the von K\'arm\'an mean flow cooperate to generate a self-sustained dynamo. 

The existence of a kinematic dynamo threshold for the VK flows, together with the possibility to bring its value within experimental reach (in terms of power requirements) has motivated the sodium experiments in Maryland, Cadarache and Wisconsin~(\cite{VKSGydro} and references therein). However, it has also been recognized early~\cite{MordantGlobal} that VK flows at high Reynolds numbers have strong fluctuations. As a result the instantaneous mean flow structure can differ significantly from the time-averaged flow. This has motivated investigations 
regarding the influence of noise on the dynamo bifurcation. 
For instance, it has been proposed that the VK flow may lead to an intermittent dynamo~\cite{LathropOnOff,Wisconsin,Leprovost}:  burst of dynamo activity would occur, triggered by transient flow structures that are most efficient at generating magnetic induction. 

Induction measurements in the presence of an externally applied field have been made in the VKS experiment~\cite{VKSPoF}. They have demonstrated the existence of fluctuations in the magnetic induction due to small scale turbulence and also generated from the instationarity  of the large scales. For example, it has been observed that the local fluctuations of induction measured during a time interval $\tau$ increase  logarithmically with $\tau$~\cite{VKSPoF}. This finding can be linked to the existence of a $1/f$ scaling in the low frequency domain of the magnetic induction spectrum, a feature that has been also observed in numerical models~\cite{PontyPRL1} and in the dynamo regime of the Karlsruhe experiment~\cite{KarlsruheRecent}.

The purpose of this article is to explore further the above ideas. We relate the fluctuations in time of magnetic induction to large scale fluctuations in the geometry of VK flows. We use a VK flow of liquid Gallium, and we study the magnetic induction in the presence of an externally applied field. Besides being relatively easy to handle in the laboratory, the magnetic Reynolds number in Gallium flows reaches values of order 1: induction effects are measurable and the non-linearities are not yet strong. The amplitude of the induced field is often an order of magnitude smaller than the amplitude of the applied field, and, in the quasi-static limit~\cite{bourgoinPoF}, the induction equation (\ref{eqB}) is reduced to $\nabla \times \left( \langle \bU \rangle  \times \bB_0 \right) + \lambda \Delta \bB = 0$, with $\bB_0$ the applied field. When it is uniform in space, the induced field solves $\lambda \Delta \bB = - (\bB_0 \cdot \nabla) \langle \bU \rangle$ so that it is directly related to the velocity gradients in the flow (although the solution is not local;  the overall topology of the velocity gradients and the boundary conditions do enter in the solution of this Poisson equation). 

The induction measurements are made using a new probe made of a line of magnetic field sensors. It samples simultaneously one component of the magnetic field at several location within the flow. The paper is organized as follows: in the next section, we describe the experimental set-up and we characterize our multi-probe magnetic measurements. The results are detailed in section III and their implications regarding dynamo action in VK flows are discussed in  section IV.

\section{Set-up and measurements}

\label{exp_setup}
\subsection{Flow.}
Our experiments are carried out in the setup sketched in Fig.\ref{setup}. The flow is produced by the rotation of one or two disks inside a stainless steel cylindrical vessel filled with liquid Gallium. The cylinder radius  $R$ is 97~mm and its length is 323~mm. The disks have a diameter equal to 165~mm and are fitted a set of 8 blades  with height $10$~mm. They are separated by a distance $H =203$~mm. The disks are driven by two 11~kW AC-motors  which provide a constant rotation rate in the interval $\Omega \in [0.5, 25]$~Hz with a stability of about 0.1\%.  The system  is cooled by a set of coils located behind the driving disks; the experiments are made with the flow kept at a temperature in an interval between $42^{\circ}$C and $48^{\circ}$C. The fluid is liquid Gallium  (density $\rho = 6.09 \times 10^{3} \; {\rm kg} \; {\rm m}^{-3}$) whose electrical conductivity is 
 $\sigma = 3.68 \times 10^{6} \; {\rm ohm}^{-1} \; {\rm m}^{-1}$. Its kinematic viscosity is  $\nu = 3.1 \times 10^{-7} \; {\rm m}^{2} \; {\rm s}^{-1}$.  The integral kinematic and magnetic Reynolds numbers of the flow are defined as ${\rm Re} = {2\pi R^{2}\Omega}/{\nu}$ and ${\rm R_m} = 2\pi \mu_{0}\sigma R^{2}\Omega$. Values of ${\rm R_m}$ up to $5$ are achieved, with corresponding ${\rm Re}$ in excess of $10^6$.

\subsection{Induction measurements.}
\begin{figure}[htb]
\centerline{\includegraphics[width=14cm]{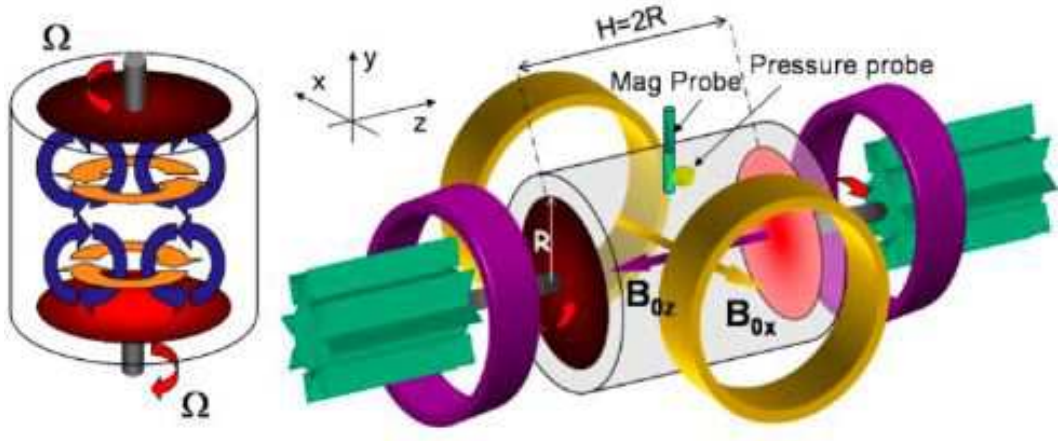}}
\caption{Set-up : (a) schematics of the von K\'arm\'an mean (time-averaged) flow geometry for counter rotating disks at equal rate $\Omega$. The rotation is opposite in each  half of the cylinder, leading to a strong differential rotation and the recirculation loops in the flow create a stagnation point: the poloidal flow converges in the mid-plane and is drawn towards the disks on either side. (b) experimental arrangement: coils and magnetic measurement probe.} 
\label{setup}
\end{figure}

Two sets of coils (see Fig. \ref{setup}) are used to apply an external magnetic field $\bB_0$, either parallel or perpendicular to the axis of the cylinder. The locations of the coils is such that the configuration is close, but not strictly equal, to a Helmholtz geometry. Variations of applied field intensity over the flow volume are of the order of 18\% of the mean, for axial or transverse applied fields. Its magnitude $B_0$ is less than 100~G, so that the interaction parameter,
\begin{center}
 $N = \sigma B_0^{2} L / \rho U = \sigma B_0^{2} 2R / \rho 2\pi R \Omega \sim 10^{-3}$ \ ,
\end{center}
is quite small. One can safely neglect the back-reaction of the Lorentz forces on the velocity field. 

Magnetic induction measurements are performed using a multi-sensor probe. Eight Hall sensors form a linear array which can be inserted inside the flow to yield  simultaneous measurements of the magnetic field along a line, {\it i.e.} measurements of magnetic {\it induction profiles}.  In the experiment, the probe array is inserted radially into the flow, in the mid-plane between the driving disks. The magnetic field is sampled at 8 locations between 1.5 and 8.5~cm from the rotation axis. The spacing between the Hall sensors is equal to 1~cm, equal to the overall diameter of the probe --  small compared to magnetic diffusive scale in the Gallium flow ($\ell_M \sim 2R/{\rm R_m}^{3/4} \geq 6$~cm).
According to the orientation of the probe, one obtains a radial profile of one component of the magnetic field,   $B_x(r_i,t)$ or $B_z(r_i,t)$ (for $i=1...8$) --- the axis are shown in Fig.~\ref{setup}. The Hall sensors are single axis elements from Sentron ($1$SA-$1$M), with a sensitivity of $0.03$ V$/$G and a frequency range from DC to $10$~kHz. We have calibrated these probes by comparison with a temperature compensated Hall probe connected to a Bell  FW-9953 gaussmeter. Their temperature dependence is small, of the order of $0.01$ G$/$K in the temperature range of interest {\it i.e.} between $40$ and $50^{\circ}$C. In typical measurements, the signals from the 8 sensors are  recorded for durations between $150$ and $480$~seconds, using a National Instrument PXI-4472 digitizer at a rate of $1000$~Hz with a 23 bits resolution.\\

\begin{figure}[hbt]
\centerline{\includegraphics[width=7cm]{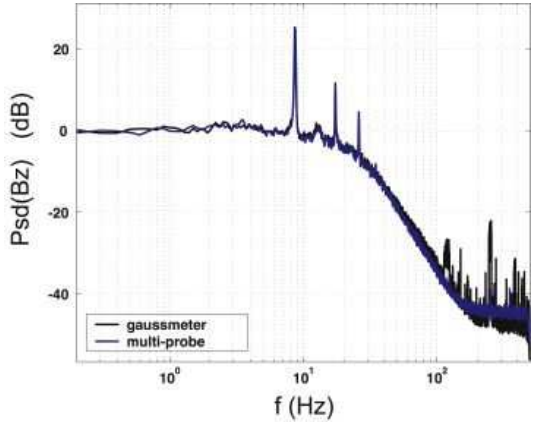}}
\caption{Spectrum obtained with one Hall sensor, compared to a measurement made using a calibrated Bell gaussmeter. Flow driven by the rotation of a single disk, at $\Omega = 10$~Hz. The main peak is at $8.5$~Hz because the flow rotation rate is lower that that of the driving disk. } 
\label{spectre1}
\end{figure}

We compare in Fig.~\ref{spectre1} a typical frequency spectrum computed using a time series from a single sensor, compared to the equivalent measurement made with the  Bell gaussmeter. In this case, the flow is set into motion by the rotation of only one of the driving disk, as in ref.~\cite{VKSalpha}; this choice is made in order to have a steady flow with little large scale fluctuations. The two curves are in excellent agreement. The dynamical range spans 80~dB. Our in-house magnetic line array has thus characteristics that are equivalent to those of the commercial single probe.

\section{Results}
\subsection{Average induction profiles}
We present here the analysis of mean induction effects which will be useful for our discussion of the fluctuations -- the reader is also referred to~\cite{bourgoinvolk}. 

We label $\bB_0$ the applied field and $\bB(\br, t)$ the field induced by the flow motion. We write the total magnetic field as $\bB_0 + \langle \bB \rangle + \bb$, where $\langle \bB \rangle $ is the time average of the induced field, and $\bb$ the fluctuating part ($\langle \bb \rangle = 0)$. By definition, the $k$-th component of time-averaged induced field at location ${\mathbf r}_j$ is
\begin{equation}
\langle B_k(r_j) \rangle = \frac{1}{T} \int_0^{T} dt \, B_k(r_j, t) \ . 
\end{equation}
In practice we have chosen $T \sim 100 \; \Omega^{-1}$, where $\Omega^{-1}$ is the period of rotation of the driving disks. 

\begin{figure}[hbt]
\centerline{\includegraphics[width=14cm]{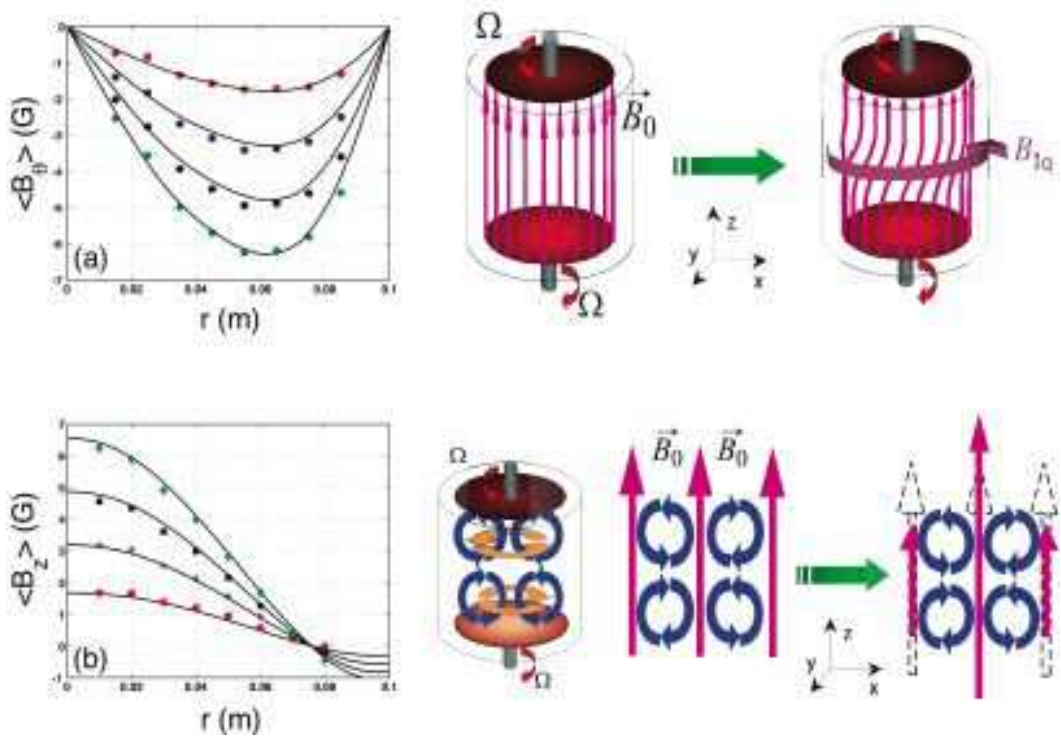}}
\caption{(Top) Time-averaged induction profile $\langle B_\theta \rangle(r_j)$ for an axial applied field $B_{0,z}$. (left): measurements for ${\rm R_m} = 1, 2, 3, 4$, top to bottom (the symbols are the measurements and the solid line is the result of the numerical simulation based on the time-averaged flow). (right): schematics of the Omega effect responsible for the induction.\\
(Bottom): corresponding figures for $\langle B_z \rangle(r_j)$ measurements in the case of a transverse applied field $B_{0,x}$. In this case the induction is due to the stretching and compression of the field lines by the axial flow. } 
\label{meanprofiles}
\end{figure}

In Fig.~3(a,b) we show the induction profiles $\langle B_\theta \rangle(r_j)$ and $\langle B_z \rangle(r_j)$ measured when an external field $\bB_0 = B_{0} \hat{z}$ is applied, directed along the axis of the cylinder. As detailed in previous studies (e.g.~\cite{bourgoinvolk}), the azimuthal component  $\langle B_\theta \rangle$ is mainly generated by the counter rotation induced by the driving impellers on each side of the mid-plane, {\it via} the Omega effect~\cite{Moffat} 
\begin{equation}
\lambda (\Delta \langle \bB \rangle)_\theta = - B_{0,z} \partial_z \langle U_\theta \rangle \ .
\end{equation}
On the other hand, when one probes the induced field along the axis, the main effect is due to the axial stretching cause by the motion of the fluid towards each driving disks (the stretching component of the stagnation point in the mid plane). 
\begin{equation}
\lambda \Delta \langle B_z \rangle = - B_{0,z} \partial_z \langle U_z \rangle \ .
\end{equation}
These mechanisms are sketched in figure~3. We stress that these profiles are obtained from simultaneous sampling in space of the magnetic field, and not from a series of independant measurements in which a single probe is successively located at a set of radial positions, as previously reported~\cite{VKSPoF,bourgoinvolk}.  

We then compare the measurements to a numerical solution of the stationary induction equation:
\begin{equation}
0 = \nabla \times \left( \langle \bU \rangle \times ( \bB_0 + \langle \bB \rangle ) \right) + \lambda \Delta \langle \bB \rangle  \ .
\label{eqIndSta}
\end{equation}
In the numerical computation, the time-averaged velocity field is taken from experimental measurements~\cite{Marie} and the induction equation is solved using the iterative method described in detail in~\cite{bourgoinPoF}  (the scheme is based on finite difference computations, and thus allows for the implementation of realistic  electromagnetic boundary conditions). Because ${\rm R_m}$ remains small (less than 5), we have found that in each case it was sufficient to compute the induction up to order 3 in ${\rm R_m}$. In Fig.~3, the profiles obtained numerically are shown as solid lines. One observes an excellent agreement with the experimental measurements. 
It indicates that the mean induction is very well accounted for by the time-averaged flow --- equation~(\ref{eqIndSta}).  

Using all combinations for the directions of the applied and measured induced magnetic fields, we have observed that the induction is dominated by two main mechanisms: the differential rotation generated by the counter rotating disks (the toroidal part of the flow) and the existence of a stagnation point in the mid plane (the poloidal component).  For an applied field parallel to the rotation $z$-axis (case in Fig.~3), the toroidal velocity gradients induce a toroidal magnetic field (the Omega effect) and the poloidal velocity gradients generate a stretching of the applied field lines. When the field is applied transverse to the rotation axis (e.g. along the $x$-axis) the poloidal flow produce a compression of the applied field lines while the toroidal flow generates a $z$-component associated to the connection of the magnetic field lines induced along the $y$-axis on each side of the mid plane. This effect is quite sensitive to the electromagnetic boundary conditions; we have called it the BC-effect~\cite{bourgoinPoF}. 

Regarding the mean (time-averaged) value of the magnetic induction, our main finding is  that the average induction coincides with the induction predicted from the average flow --- at least for the magnetic Reynolds numbers reached in this Gallium experiment. One has
\begin{equation}
\langle \bB \rangle = {\rm induction \; from \; } \langle \bU \rangle \ ,
\end{equation}
{\it i.e.}, the mean induced field effectively solves $ \nabla \times \left( \langle \bU \rangle \times \bB_0 ) \right) + \lambda \Delta \langle \bB \rangle = 0$. We will report elsewhere a detailed analysis of the induction effects due to the turbulent small scale  fluctuations~\cite{VolkPerm}; its main finding is to confirm the above statement: the small scale turbulent fluctuations do not contribute to the mean induction.

\subsection{Fluctuations}
\subsubsection{General considerations}
As pointed out in the introduction, due to the very small value of the Prandtl number,  the flow is fully turbulent. The Reynolds number is larger than $10^6$, even for the relatively moderate values of the magnetic Reynolds number reached here. The local velocity fluctuations reach 35\% of the mean speed, and so do the induced fields, as shown in figure 4(a,b). The fluctuations are large and occur over a very broad range of time scales. The fastest fluctuations are associated with the shearing of the applied field by the turbulent small scale motion~\cite{Odier98}.  Komogorov's scaling for the inertial range of motions together with the use of Taylor's hypothesis predict an $f^{-11/3}$ for frequencies larger than that, $\Omega$, of the forcing. This is due to the fact that  in our experiment ${\rm R_m}$ is of order one, so that the magnetic dissipation scale $\ell_M$ is of the order of the flow size. Our measurements -- figure 4(c) -- show a steeper slope, of the order of $-4.6$ for driving disks fitted with blades. The Kolmogorov prediction $-11/3$ was observed in the flow generated by flat rugose disks~\cite{Odier98}.  The slope of the spectra in the high-frequency (dissipative) region does not depend upon which component is being probed, nor on the rotation rate of the disks. 

Another observation is that the magnetic induction also fluctuates in a broad range of long time scales, as indicated by the behavior of the power spectra for frequencies lower than the disks rotation rate $\Omega$ -- Fig.~4(c).  In addition, it is clear  in figure~4(c) that the lower end of the frequency spectra does depend on which induction mechanism is at work. In the case of an axial applied field, the fluctuations in the toroidal  component of the induction are related to fluctuations in the differential rotation. The low frequency part of the spectrum is close to an $f^{-0.5}$ behavior. The fluctuations of the induced field in the direction of a transverse applied field are linked to the dynamics of the stagnation point in the mid plane. In this case the low frequency part of the spectrum has stronger fluctuations (the spectrum is close to a $f^{-1.2}$ behavior). The fluctuations of induction are associated to fluctuations in the position of the stagnation point. Concerning this slow dynamics, we have also varied the disks rotation rate between 5~Hz and 20~Hz without detecting any noticeable change in the above features; the spectra as in figure~4(c) collapse when rescaled by the rotation rate and the fields $rms$ amplitude. 

\begin{figure}[hb]
\centerline{\includegraphics[width=10cm]{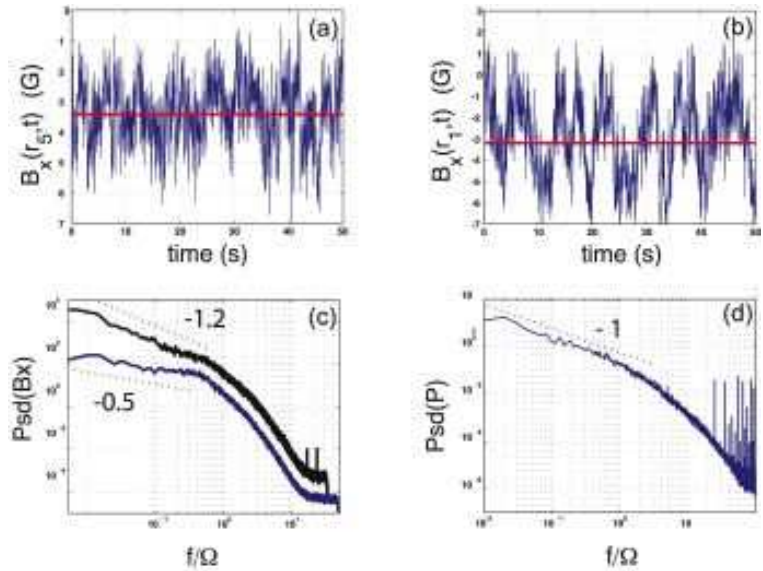}}
\caption{Fluctuations for local measurements. Counter-rotation of the disks at $\Omega = 10$~Hz. (a) $B_x$ induced component for an applied axial field $B_{0,z}$; (b) $B_x$ induced component for an applied transverse field $B_{0,x}$, (c) Corresponding time spectra (black line : signal in (b); blue line : signal in (a)). The curves have been shifted vertically for clarity; (d) Power spectral density of the pressure recorded at the flow wall in the mid-plane.   } 
\label{localfluct}
\end{figure}

The slow dynamics in the induction traces back to the evolution of the hydrodynamic flow. This is evidenced by measurements of pressure fluctuations at the wall; one observes a $1/f$ behavior  in the same range of frequencies where the magnetic fields shows a long-time dynamics -- figure~4(d). However, the magnetic measurement is quite sensitive because is probes the velocity gradients selectively. The poloidal and toroidal components are independantly probed by chosing the direction of the applied field and the particular component of the induced field under inspection.

\subsubsection{Profile fluctuations}
We analyse below the fluctuations of induction `profiles' which we define as the set of measured magnetic field values $\{B_k(r_1, t), ... B_k(r_8, t)\}$ with $(r_1, ... ,r_8)$ radial positions at distances between 1.5 and 8.5~cm from the axis of rotation, and the applied field along direction $\hat{\imath}$. The choice of the $(i,k)$ couple determines which component (toroidal or poloidal) of the flow is probed. We stress again that we do not address here the question of the contribution of the small scales of turbulence. The recorded signals $B_k(r_n, t)$ are low-pass filtered with a corner frequency equal to 3 times the disks rotation rate $\Omega$.

\begin{figure}[hbt]
\centerline{\includegraphics[width=9cm]{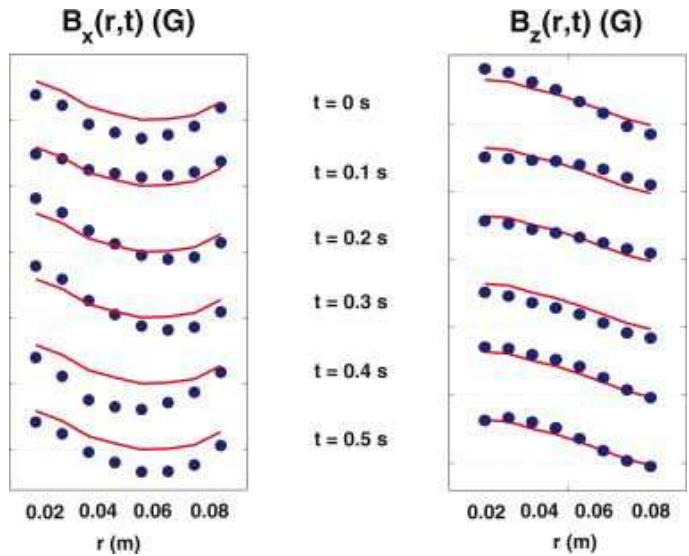}}
\caption{Examples of fluctuations of profiles.  Counter rotation of the disks at $\Omega=10$~Hz; applied field along the $z$-axis, $B_{0,z} = 24$~G. (left) transverse induced field due to differential rotation; (right) axial induced field due to the stretching effect.} 
\label{profiles}
\end{figure}

 Examples of the evolution in time of such profiles are shown in figure~5. The profiles are shown at time intervals equal to the period of rotation of the disks, and one observes significant fluctuations with respect to the time-averaged curves.  In addition, the points in the profiles tend to vary as a block.  This is evidenced by computing the correlation function, between, say the first element in the array, and the progressively more distant ones. The resulting variation is shown in figure 6(a), where one observes that the maxima of the correlation functions decrease very slowly from 0.9 for the second element down to about 0.4 for the last one. These values do not change as the disks rotation rates are varied from 5~Hz to 20~Hz. For an estimation of a typical correlation length, one notes that the curve can be fitted by an exponential function $\exp{-(r/r_0)}$, with $r_0 \sim R$, the radius of the cylindrical vessel.  This confirms that the fluctuations in the induction are a global feature rather than a local one. In addition, we have observed that the correlation functions are symmetric and peaked around the zero time-lag value.  This behavior is also evidenced in figure 6(b) which shows a space-time diagram of the profile dynamics:  fluctuations are felt simultaneously at all points, {\it i.e.} with a response time of the order of $1/\Omega$.  

\begin{figure}[hbt]
\centerline{\includegraphics[width=6cm]{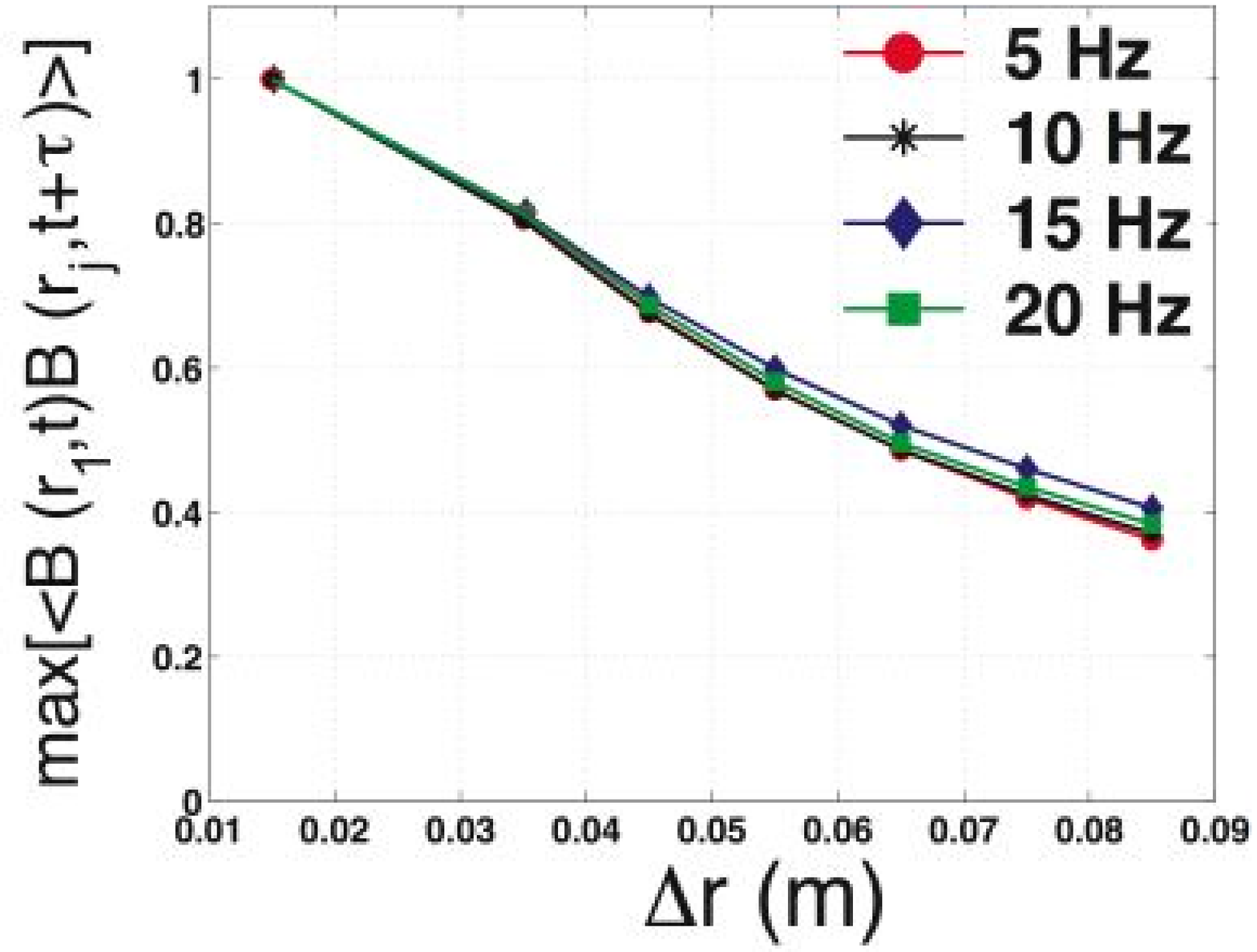}\includegraphics[width=6cm]{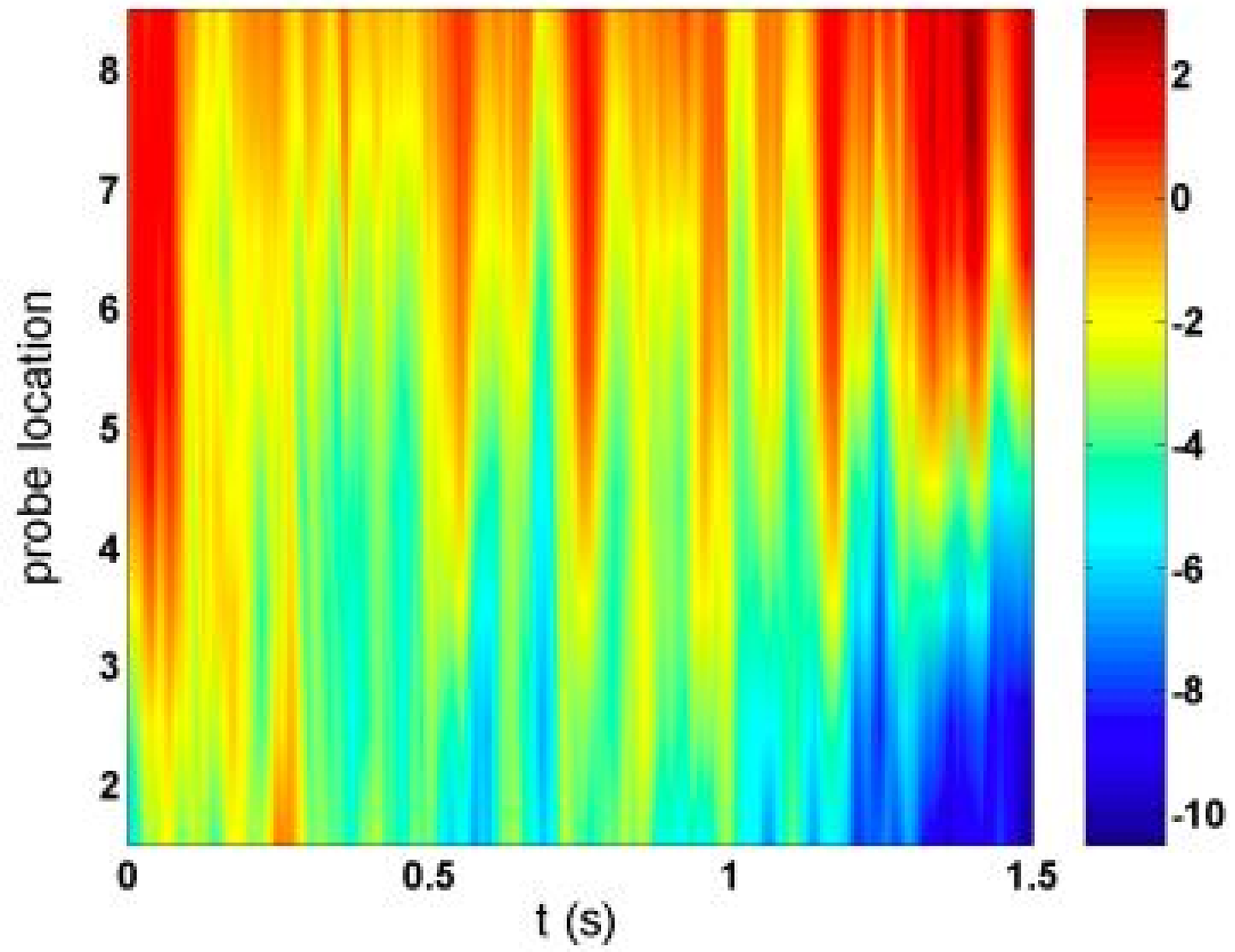}}
\caption{(a) Evolution with the distance to the first element in the array, of the maximum (reached at $\tau=0$) of the correlation function $C_{1j}(\tau)= \langle B(r_1, t)B(r_j, t+\tau)\rangle$.  Counter rotation at $\Omega = 5, 10 , 15, 20$~Hz, corresponding to ${\rm R_m}$ values between 1 and 4. (b) Space time diagram, at $\Omega = 15$~Hz. Measurements for counter rotating disks, with $B_{0,x}= 48$~G, induced field along the same direction $B_x$.} 
\label{correl}
\end{figure}

\subsubsection{Distance to the mean}
Let us compute a global distance between the mean flow induction profile and  a realisation. Using the ${\cal L}_2$ norm, we define
\begin{equation}
E_k(t) = \sqrt{\frac{1}{N} \sum_{i=1}^{N=8} \left( B_k(r_i, t) - \langle B_k(r_i) \rangle \right)^2 } \ ,
\end{equation}
We find that the time averaged value of $E$ are found to be of the same order of magnitude as the average induced field, that is $\langle E(t) \rangle \sim \langle B \rangle$. In units of the applied magnetic field, we measure $\langle E \rangle \sim 0.07 \ B_0$ for situations in which the maximum of the induced field (${\rm max}_{r_i} \langle B \rangle(r_i)$) is also of the order of $0.1 \ B_0$. These observations indicated that instantaneous induction profiles differ significantly from the time averaged computation. 
In addition, we also observe that the distance $E$ has large fluctuations away from its mean value. This is evidenced in figure 7(a) which shows the probability density functions ${\cal P}(E)$ for induction measurements probing the toroidal and poloidal flows. The curves are wide, with events that span several standard deviations. The fluctuations are found to be larger in the case of the induction due to differential rotation than when due to the stretching by the stagnation point in the mid-plane.

\begin{figure}[hbt]
\centerline{\includegraphics[width=8cm]{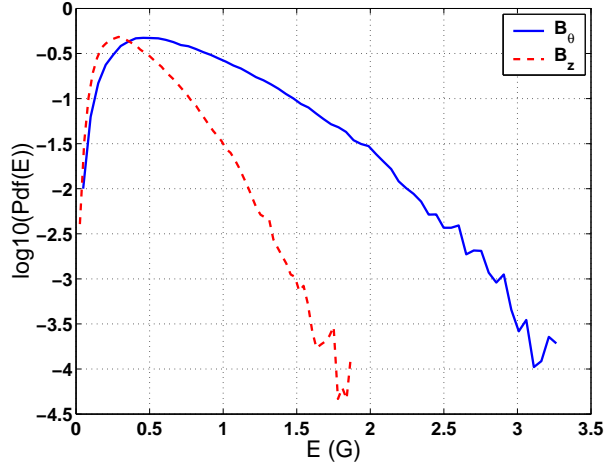}}
\caption{Probability density function of the fluctuations of the ${\cal L}_2$ distance $E(t)$ to the mean induction profile. Counter rotation at $\Omega = 10$~Hz, corresponding to ${\rm R_m} = 2$.} 
\label{distance}
\end{figure}

As in our experiment the intrinsic magnetic Reynolds numbers~(defined by the ratio ${\rm max} \langle B \rangle / B_0$~\cite{OdierBilles}) reached in our experiment is always less than 1, we believe that these fluctuations mirror the variations in the flow velocity gradients. However the geometry of these changes is lost with the use of a global ${\cal L}_2$ norm. In the next section we analyse further the profile fluctuations, with the purpose of trying to quantify how the induction deviates from the one expected from the mean flow.

\subsubsection{Polynomial analysis}
Because of the strong correlation in the signal measured by successive elements in the magnetic array, the induction profiles are smooth  and are well described by  polynomials of order three. We do not claim here that the fitting functions are the actual solutions to the induction equation. We use the polynomials as a way to take advantage of the symmetries  associated to the von K\'arm\'an flow. For instance, the mean velocity is reflection - symmetric about the mean plane, and the azimuthal velocity is zero on the rotation axis. As a result, the induction due to differential rotation in the presence of an axial applied field (the $\Omega$ effect) has a zero average in the mid plane at $r=0$. Deviations from this value can then be associated with a symmetry breaking in the mean flow pattern. We thus write an instantaneous profile as 
\begin{equation}
\{B_k(r, t)\} = a_0(t) + a_1(t) r+ a_2(t) r^2 + a_3(t) r^3
\end{equation}
and we study in the sequel the evolution of the coefficients $a_j(t)$. \\

\noindent {$\bullet$ \it Mean flow polynomials}\\
We first present the coefficients $\langle a_j \rangle$ for the mean profiles, and their evolution with the disks rotation rate. From the results presented in section 3.1 we expect that they are well accounted for by the structure of the mean flow, {\it i.e.}  $\langle a_j \rangle = a_j({\rm from} \, \langle \bU \rangle)$. 

Let us first consider the evolution of $B_x$ in the case of an axial applied field (along $z$). As detailed in section 3.1, induction in the mid plane is dominated by the twisting of the magnetic field lines by differential rotation. Thus, $\langle a_0 \rangle$ should be null because of axisymmetry and $\langle a_1 \rangle$ should increase linearly with the disk (counter) rotation rate. This is indeed observed in figure~8. We have no simple interpretation for the coefficients $a_2$ and $a_3$ which mainly ensure that the induced field vanishes at the outer cylinder (because of the insulating boundary condition). 

\begin{figure}[ht]
\centerline{\includegraphics[width=10cm]{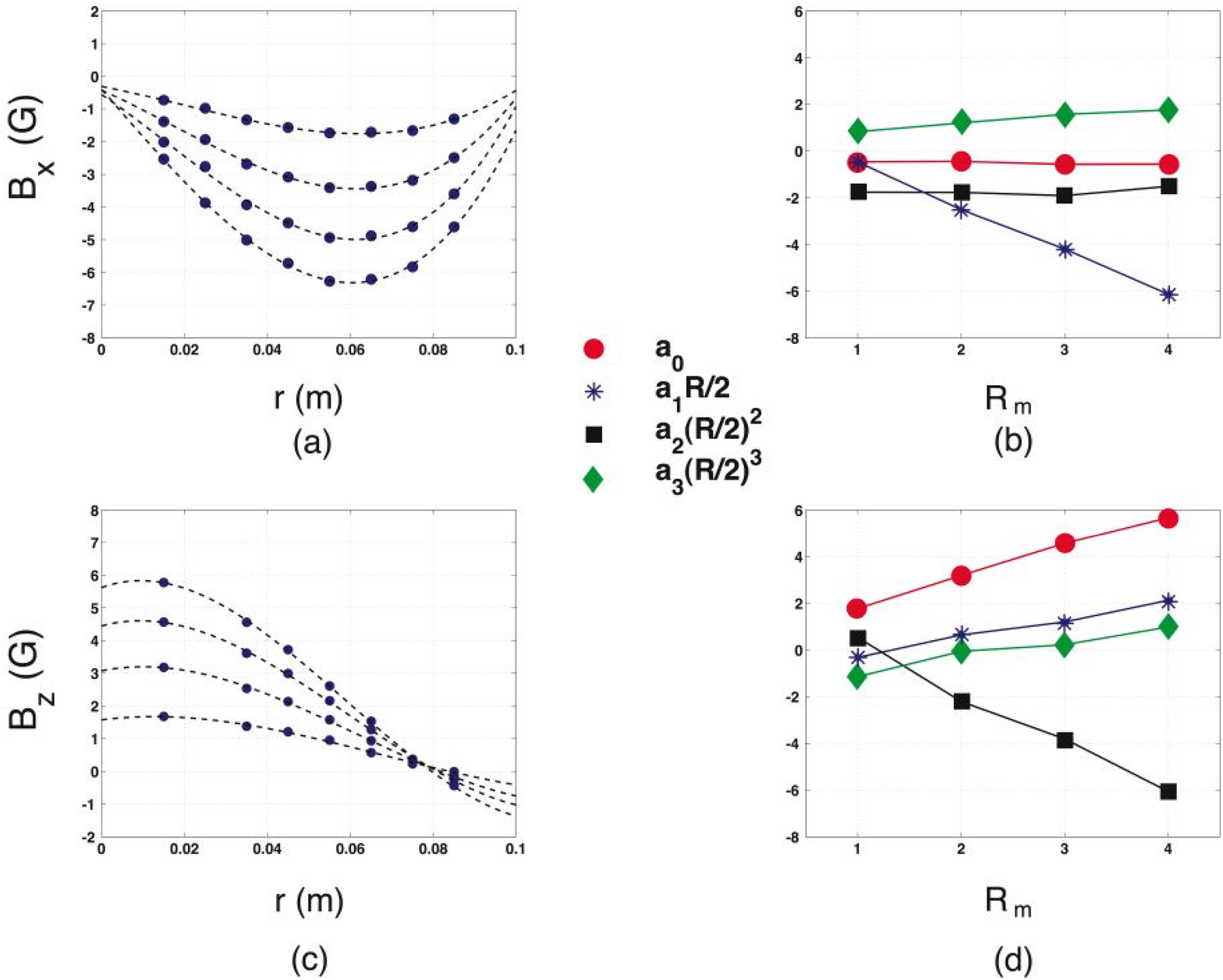}}
\caption{Polynomial coefficients for the mean induction profile. Axial applied field $B_{0,z} = 24$~G. The measured profiles and corresponding polynomial fit are for ${\rm R_m} = 1, 2, 3, 4$. (top): induction due to differential rotation; (bottom): axial stretching of the applied field.} 
\end{figure}

\begin{figure}[hb]
\centerline{\includegraphics[width=10cm]{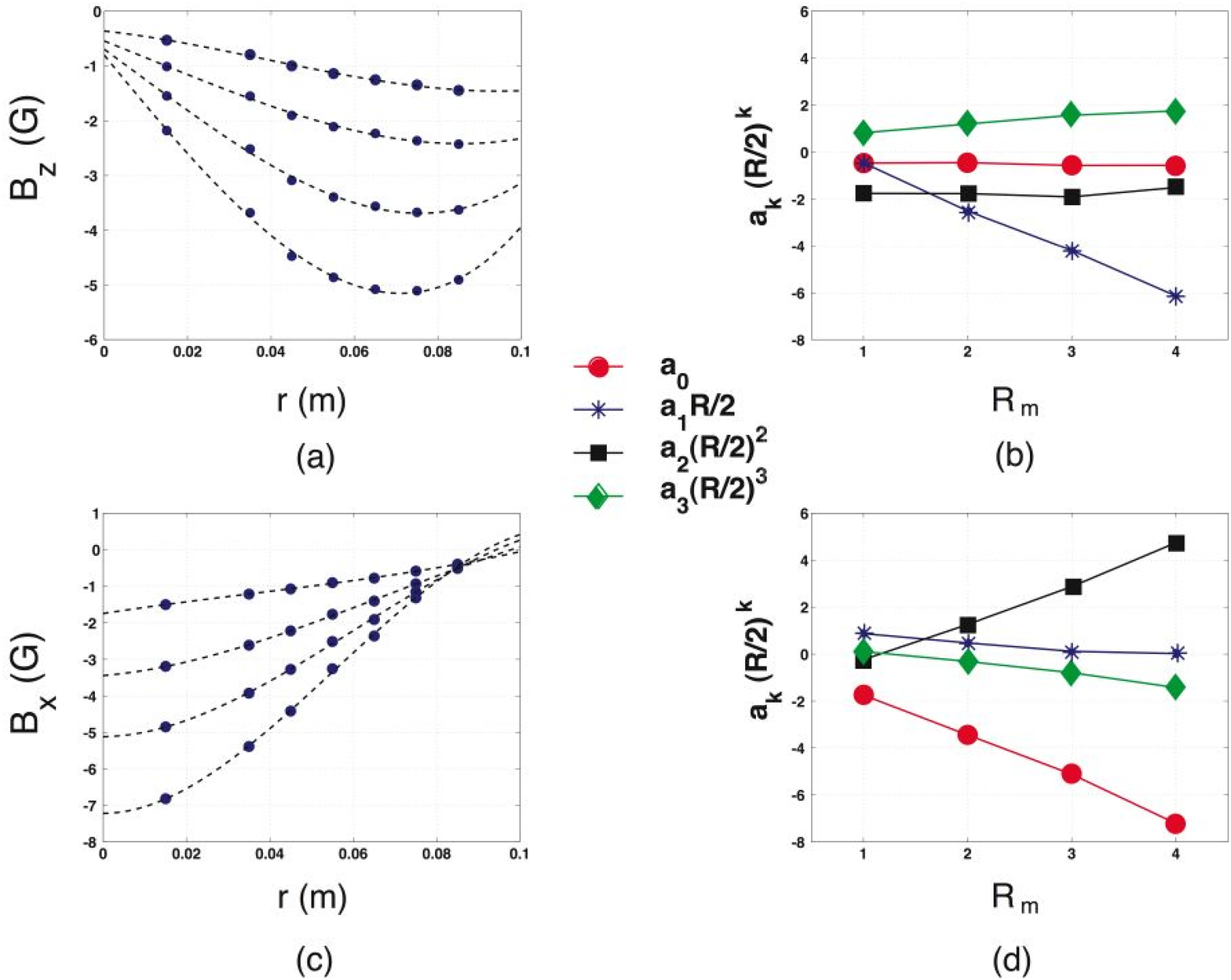}}
\caption{Polynomial coefficients for the mean induction profile. Transverse applied field, $B_{0,x} = 48$~G.  The measured profiles and corresponding polynomial fit are for ${\rm R_m} = 1, 2, 3, 4$. (top): induction due differential rotation; (bottom): transverse compression of the applied field in the mid-plane.} 
\end{figure}
\clearpage

Similarly, when ones probes the $z$-induced field  the main effect is due to the stretching of the applied field lines. Since the applied field is uniform, one expects a non zero component $\langle a_0 \rangle$, which increases with  the disk (counter) rotation rate -- as confirmed in figure~8(c). The other contribution is in the $a_2$ coefficient, since here the axisymmetry requires that the odd terms be null. 

The situation is equivalent when the applied field is transverse ${\mathbf B}_0 = B_0 \hat{x}$. The average field induced along $x$ in the mid plane comes from the compression of the applied field by the converging flow. The dominant contribution is in $a_0$ (negative) which varies linearly with the rotation rate -- figure~9(d) -- while the axisymmetry imposes $\langle a_1 \rangle  = \langle a_3 \rangle = 0$.  When one probes $B_z$, the dominant contribution comes from the differential rotation whose effect is not felt on the rotation axis, so that $\langle a_0 \rangle = 0$. The dominant contribution is in $\langle a_1 \rangle$, which varies linearly with $\Omega$ (figure~9(b)).\\

\noindent {$\bullet$ \it Fluctuations of the polynomial coefficients}\\
Let us first discuss the case of the fluctuations in the Omega effect, {\it i.e.} for measurements of the toroidal field induced by differential rotation when the applied field is axial. The time variations of the leading coefficient $a_1$ are shown in figure~10(a). One observes very strong fluctuations, although the probability density function is quasi-Gaussian -- Fig.~10(b).  One finds  $a_{1,{\rm rms}} / \langle a_1 \rangle = 114\%$ at a (counter) rotation rate $\Omega=10$~Hz. The spectrum in figure~10(c) indicates that the fluctuations in the profile have a long-time behavior (the slope of the spectrum in the low frequency range is close to -0.7). One thus observes that the fluctuations in the induction profile correspond to a slow process, compared to the turbulence fast small scales and also compared to the disk turn-over time. The fact that the entire induction profile changes in time is clear in figures~10(d,e) which show the correlation between the coefficients $a_0$ and $a_1$. One observes that $a_0$ and $a_1$ are anti-correlated, with $a_0 \sim 0$ only for small deviations of $a_1$ about its mean. In most configurations $a_0$ is non-zero.   As at small ${\rm R_m}$ the induction mirrors the evolution of the velocity, we conclude that the  flow has a slow dynamics with strong deviations from the mean von K\'arm\'an geometry. 

\begin{figure}[hbt]
\centerline{\includegraphics[width=14cm]{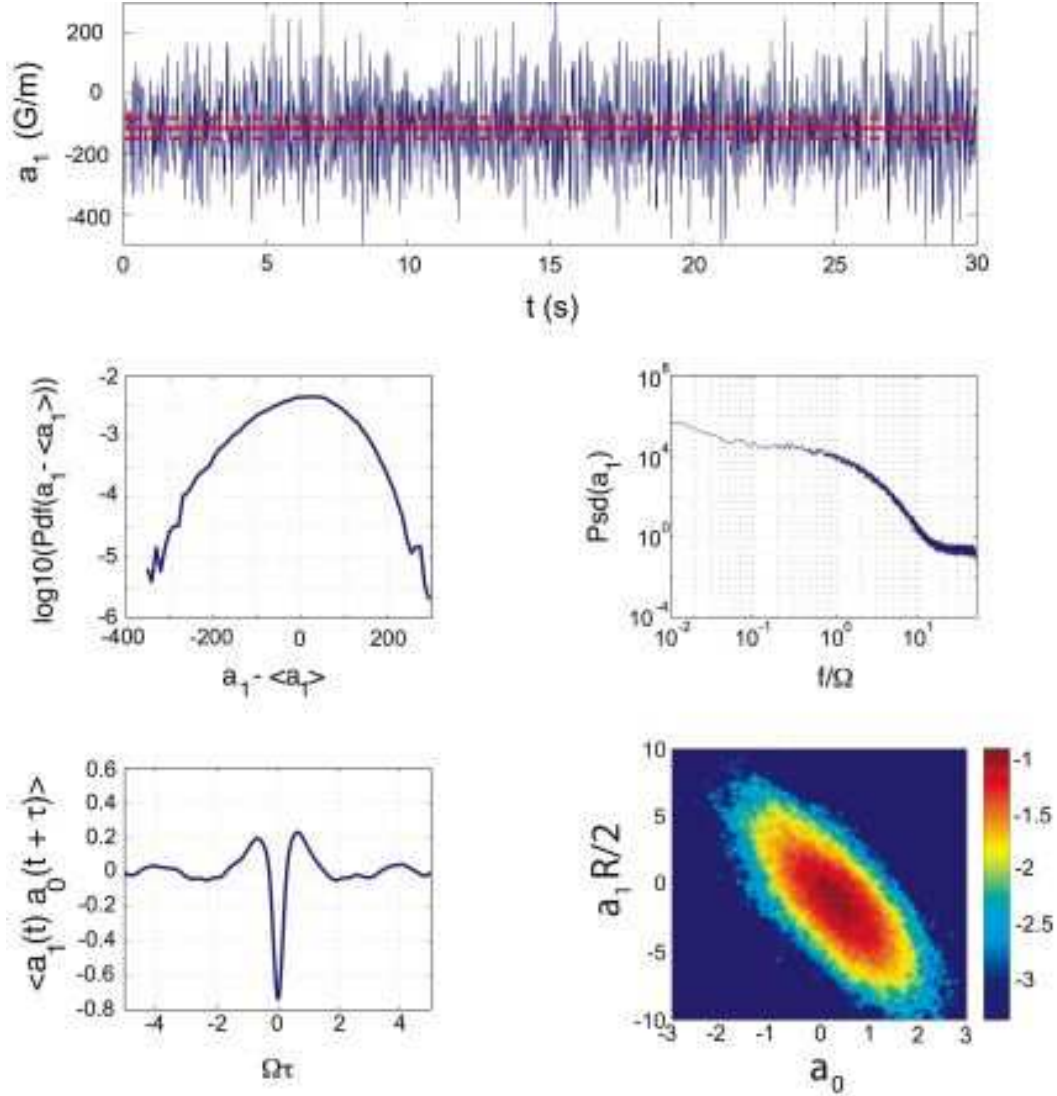}}
\caption{Fluctuations of the polynomial coefficients, for the Omega effect: a magnetic field $B_{0,z} = 24$~G is applied along the rotation axis, and the induction profile $B_\theta \equiv B_x$ is measured. Counter-rotation of the disks at $\Omega = 10$~Hz. (a) time evolution of the polynamial coefficient of order 1; (b) corresponding centered probability density function; (c) time spectrum; (d) and (e): cross-correlation and joint probability density function for the coefficients $a_0(t)$ and $a_1(t)$.} 
\end{figure}

We now compare the fluctuations in time for the leading polynomial coefficient and each choice of $(B_{0,i}, B_j)$ -- figure~11. Two kinds of behavior are evidenced. When $(i,j) = (z,x)$ or  $(i,j) = (x,z)$, {\it i.e.} as one probes the toroidal flow (influence of differential rotation) the fluctuations are large. When the induction probes the poloidal component (stagnation point), {\it i.e.} for $(i,j) = (z,z)$ or  $(i,j) = (x,x)$,  the fluctuations are reduced by a factor 3. Note that in figure~10 the coefficients are compared in gauss, so that the $a_1$ values have been multiplied by a length chosen as half the cylinder radius. This choice is justified because in the profiles the induction is maximum at $R/2$ in the corresponding case. In addition, $R/2$ is also close to the  diffusive length. Another feature is the change in the ratio of the $rms$ fluctuation amplitude to the mean. In the case of an applied field parallel to the rotation axis, we had $a_{1,{\rm rms}} / \langle a_1 \rangle = 114\%$ at a (counter) rotation rate $\Omega=10$~Hz for induction resulting from the differential rotation. For the induction due to the pumping motion towards each disks, one has $a_{0,{\rm rms}} / \langle a_0 \rangle = 20\%$. Note that a $20\%$ fluctuation level is what one observes for induction effects in the case of a single rotating disk~\cite{bourgoinvolk}. One thus finds that fluctuations in the induction are larger when associated to the toroidal flow compared to the poloidal velocity component. This is confirmed in figure~11(c,d) where the applied field is transverse to the rotation axis. There $a_0$ for the compression effect fluctuates with a level  $a_{0,{\rm rms}} / \langle a_0 \rangle = 50\%$, while $a_1$ for the BC-effect has $a_{1,{\rm rms}} / \langle a_1 \rangle = 160\%$. In all cases, the PDFs of the time fluctuations of the coefficients are quasi-Gaussian.  \\

\begin{figure}[hbt]
\centerline{\includegraphics[width=8cm]{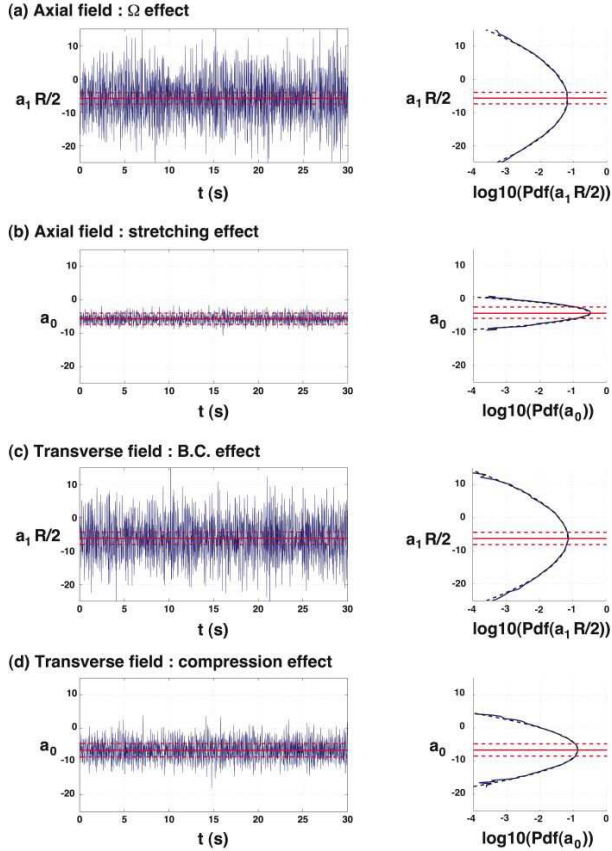}}
\caption{Fluctuations of the polynomial coefficients, for four orientations of the applied and measured components, $(B_{0,i}, B_j)$.  (a) $(i,j) = (z,x)$; (b) $(i,j) = (z,z)$; (c) $(i,j) = (x,z)$; (d) $(i,j) = (x,x)$. Counter-rotation at $\Omega = 10$~Hz. The solid red line corresponds to the mean value and the dashed line to a level of fluctuation equal to 20\% of the mean. 
} 
\end{figure}

\noindent {$\bullet$ \it Very slow quasi-periodic modes}\\
We must mention that in figure~11(d), the profiles were high-pass filtered at a frequency above 3~Hz (for a disk rotation rate of 10~Hz) before processing . The reason is that there is a quasi-periodic evolution in the induction profile, as can be observed in the  signal displayed in figure~12 (and previously in Figure~4(b)).  The unfiltered  time  variations of  $a_0(t)$  are shown in figure~12(a), and its corresponding time spectrum and probability density function in figures~12(b,c). The spectrum has a marked peak at a frequency $f = 0.1 \; {\rm Hz} \sim \Omega / 100$, with the quasi-periodic evolution also reflected in the bimodal shape of the probability density function.  

\begin{figure}[hbt]
\centerline{\includegraphics[width=8cm]{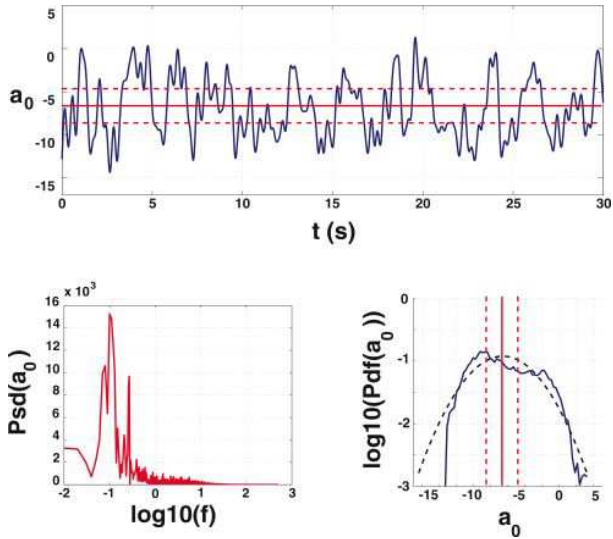}}
\caption{Transverse field and compression effect : applied field $B_{0,x}$ and induced component $B_{x}$. (a) time variation of the leading polynomial coefficient $a_0$ (low-pass filtered at frequencies lower than 3~Hz); (b, c) time spectrum and probability density function of $a_0(t)$. Measurement with counter-rotating disks at $\Omega = 10$~Hz.} 
\end{figure} 

In this configuration, the external magnetic field is transverse. The induced component  probed is parallel to the applied field, and results from the compression of the applied field lines by the poloidal flow converging in the median plane. This effect strongly depends on the exact location of the shear layer~\cite{bourgoinvolk}. Its oscillations with respect to the mid-plane would generate oscillations as in Fig.~12(a). We note that these slow oscillations are reminiscent of the global instabilities of the median shear layer, previously discovered and studied in water prototypes by CEA-Saclay group~\cite{MarieThesis,RaveletPoF}. \\

\clearpage
\section{Discussion and concluding remarks}
In the last case discussed above, the fluctuations are attributed to the existence of large scale (vortical) structures, possibly due to the roll-up of the shear layer in the von K\'arm\'an flow. When these structures exists, it is obvious that the flow is no longer in the averaged configuration pictured in figure~1(a). We would like to show now, that even if one discards the influence of these large scale coherent structures (as we have done by high-pass filtering the data), the inherent turbulence of high Reynolds number von K\'arm\'an flows is such that the fluctuations in the instantaneous flow geometry are very large. The instantaneous flow configurations differ from the mean flow, not only in regards to the amplitude of the toroidal and poloidal velocity components, but also in regards to the symmetries and to the flow overall structure. 

We first comment again about the fluctuations, and we study in greater details the  induction profile for the Omega effect. In this case the induction is less sensitive to displacements of the mid-plane shear layer because the differential rotation is found to be rather uniform in the center of the flow~\cite{bourgoinvolk}. For $(B_{0,z}, B_{x})$, the $a_0$ coefficient should vanish because of axisymmetry. Its actual fluctuations in time are shown in figure~13. The mean value $\langle a_0 \rangle$ is equal to $0.5$~gauss; although one expects $\langle a_0 \rangle = 0$, the recorded mean value is really within the precision of our polynomial analysis (limited by the number of elements in the array probe).  On the other hand, the $rms$ amplitude of fluctuations for $a_0$ is $a_{0, rms} = 1.25$~G (for an applied field equal to  24~G), well above its mean. This is a significant variation, even when compared to the mean induction ($\langle a_1 \rangle R/2 = -2.5$~gauss). \\

\begin{figure}[hbt]
\centerline{\includegraphics[width=14cm]{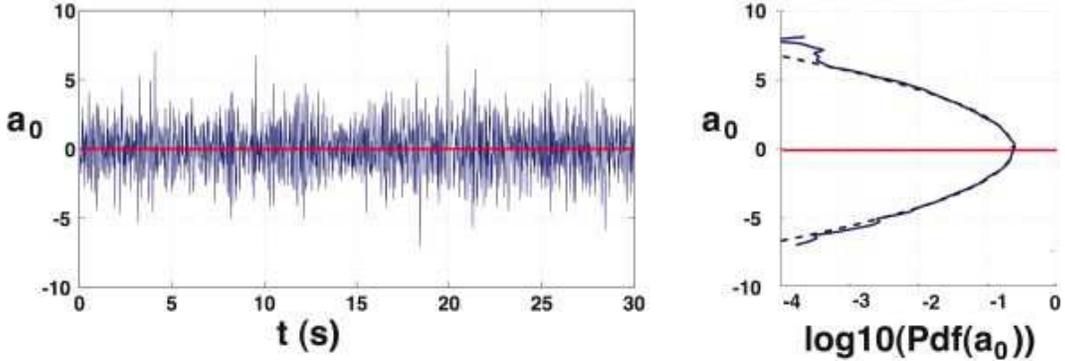}}
\caption{$\Omega$ effect, for a counter rotation at $10$~Hz. Time fluctuations of the $a_0$ coefficient (shifted by 0.5~G (see text) so that $\langle a_0 \rangle = 0$ in the plot).} 
\end{figure} 

Let us now discuss changes in the flow geometry. We compute how long the flow remains in the neighbourhood of its mean configuration, by imposing that both $a_1$ and $a_0$ take values within a standard deviation of their mean. In this case, one finds again that no longer than 50\% of the time is spend in the mean von K\'arm\'an geometry. We stress that: (i) a similar conclusion is reached if one does the estimation from induction effects generated by the presence of the stagnation point (stretching of the axially applied field lines or compression of a transverse field); (ii) the large changes in the polynomial coefficients are related to the long-time dynamics of the flow ({\it i.e.} to frequencies lower than that of energy injection into the flow).

At the low magnetic Reynolds numbers probed in this Gallium experiment, the above effects can be attributed to the hydrodynamics of the flow. Note that such is not necessarily the case for sodium experiments in which non-linear induction processes~\cite{VKSGydro,VKSalpha} or dynamo~\cite{Riga,Karlsruhe} take place. The understanding of such long time dynamics in confined turbulent flows is a challenge. It has now been reported in several experiments (e.g. recently in Rayleigh-B\'enard convection~\cite{Castaing}) but its understanding is still elusive. For instance no known argument gives the time scale of the slowest motion. 

These slow global changes in the geometry of the flow may not favor dynamo action. First because all configurations may not be consistent with self generation -- in the sense that the velocity field $\bU(\br, t \; {\rm fixed})$ may not lead to a positive growth rate when inserted in the induction equation. Second, in order for the the magnetic field to grow, the flow must be maintained for many kinetic advection times; indeed one has for the magnetic diffusion time $\tau_M = R^2/\lambda \sim {\rm R_m} \tau_{NL}$, with $\tau_{NL} = R/U$ the time scale of the forcing. Altogether these arguments indicate that a stable flow configuration is desirable for the self-generation of a stationary dynamo. For instance, for flows generated by the rotation of only one disk with the other kept at rest, we have observed that the fluctuations are much less --- cf. Figure~14. The flow spends most of its time in the $s_1t_1$ configuration imposed on average by the driving. As a result, depending on the forcing and large scale hydrodynamic evolution of the flow, one may have to be cautious with approaches that estimate the dynamo threshold from mean flow geometries. This procedure may be valid for purely helical flows in which we have not detected strong fluctuations about the mean; it certainly did yield a correct estimate of the onset of the Riga dynamo~\cite{Riga}. However, it may not be the case for the flow generated by counter-rotation in the von K\'arm\'an geometry. The observed slow dynamics is associated with important changes in the flow topology, and a mean field kinematic simulation may underestimate the threshold. \\

\begin{figure}[hbt]
\centerline{\includegraphics[width=12cm]{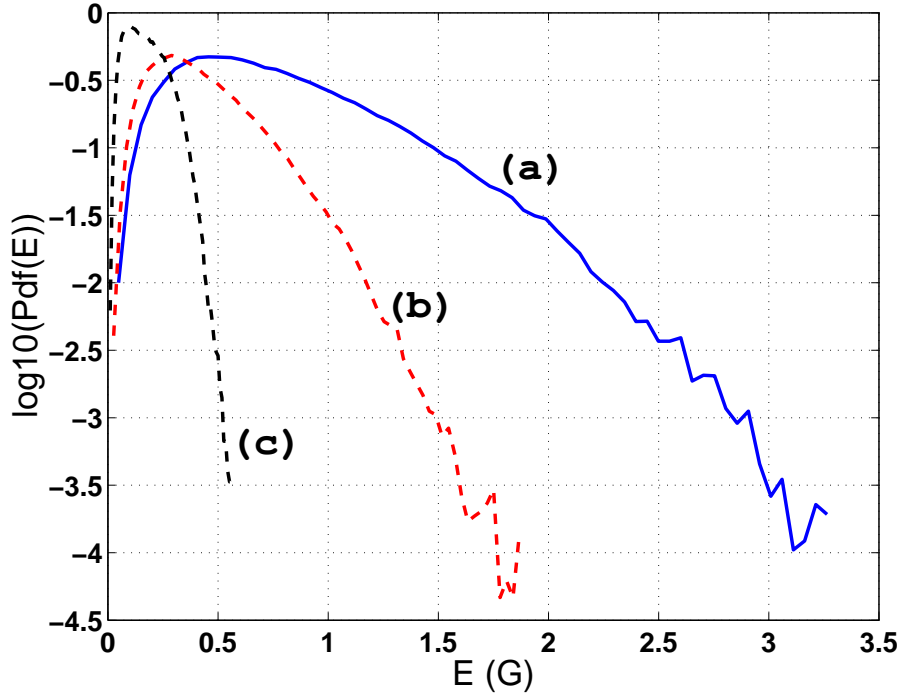}}
\caption{Probability density function of the fluctuations of the ${\cal L}_2$ distance $E(t)$ to the mean induction profile. (a,b) flow generated by counter rotating disks at $\Omega = 10$~Hz,  axial  applied field, (a) : induction due to differential rotation, (b) due to stretching; (c) flow generated by the rotation of one disk only at $\Omega = 10$~Hz, transverse applied field.} 
\end{figure} 

{\bf Acknowledgements}\\
We gratefully acknowledge many useful discussions with M. Bourgoin, P. Frick, Y. Ponty, W. L. Shew, and all the members of the VKS team (M. Berhanu, A. Chiffaudel, F. Daviaud, S. Fauve, R. Monchaux, N. Mordant, F. Ravelet). We are indebted to P. Metz and M. Moulin for technical assistance in the development of the experiment. This work is supported by the CNRS and the Rh\^one-Alpes Region Emergence program.


\end{document}